\documentclass[twocolumn]{aastex631}
\usepackage{comment}  

\newcommand{\ovi}{\ion{O}{6}}

\begin{document}

\title{Metagalactic Ultraviolet Background Tables for Calculating Diffuse Astrophysical Plasma Properties}

\author[0000-0003-3886-8326]{Elias Taira}
\affiliation{Department of Physics \& Astronomy, 567 Wilson Road, Michigan State University, East Lansing, MI 48824}

\author[0000-0001-5158-1966]{Claire Kopenhafer}
\affiliation{Institute for Cyber-Enabled Research, 567 Wilson Road, Michigan State University, East Lansing, MI 48824}

\author[0000-0002-2786-0348]{Brian W.\ O'Shea}
\affiliation{Department of Computational Mathematics, Science, \& Engineering, Michigan State University, 428 S. Shaw Lane, East Lansing, MI 48824}
\affiliation{Department of Physics \& Astronomy, 567 Wilson Road, Michigan State University, East Lansing, MI 48824}
\affiliation{Facility for Rare Isotope Beams, Michigan State University, 640 S. Shaw Lane, East Lansing, MI 48824}
\affiliation{Institute for Cyber-Enabled Research, 567 Wilson Road, Michigan State University, East Lansing, MI 48824}

\begin{abstract}
In developing a deeper understanding of the Circumgalactic Medium, one feature that is poorly understood is the nature of the ultraviolet background (UVB) and its impact on observed column densities. A wide array of UVB models have been created over the years by many different authors, each based on the latest observational data available at the time. In addition to having a large variance between model properties, the formatting between released models is also inconsistent.

This data release provides reformatted versions of several widely-used ultraviolet background  models—Faucher-Giguère et al. 2009, Haardt and Madau 2012, Puchwein et al. 2019, and Faucher-Giguère 2020—such that each model is in the same units and thus can be utilized to directly compare these models over a wide redshift range. This release also includes code to run a \texttt{cloudy\_cooling\_tools} pipeline to generate ionization tables for different UVB models.
\end{abstract}

\section{Introduction} \label{sec:intro}

To determine the ionization states of elements in the circumgalactic medium (CGM), there are two dominant mechanisms by which ionization can occur: collisional ionization and photoionization. Collisional ionization, which has a rate that scales as the square of the density, tends to be the dominant ionization mechanism in regions of high plasma density, whereas photoionization dominates in more diffuse regions as the rate scales as the magnitude of the UV background flux times the plasma density.  Given that the plasma in and around galaxies varies by many orders of magnitude in density, both of these phenomena are relevant.

While collisional ionization is well-understood from N-body, particle-in-cell, and hydrodynamical simulations, photoionization relies on the intensity of incident ionizing radiation on the CGM. This radiation is often assumed to be exclusively from a metagalactic ultraviolet background (UVB), a roughly uniform radiation field that is produced by the emission of massive stars, post-main-sequence stellar populations, and quasars. This can be quite difficult to model as determining this radiation requires significant assumptions about the radiation produced by the populations in question, wavelength-dependent escape fractions from galaxies, and the evolution of these phenomena over the age of the universe.  Constraining these models also requires a variety of observations of the intergalactic medium and stellar populations. As a consequence, these models have significant uncertainties associated with them.

In this data release we provide UV background fluxes as a function of photon energy and redshift for four commonly-used UVBs: \citet{fauchergiguere2009} (FG09), \citet{haart2012} (HM12), \citet{puchwein2019} (PW19) and \citet{fauchergiguere2020} (FG20).  The key contributions that we make are (1) tables that include information on all of the UVBs using consistent units for fluxes for ease of use and comparison, and (2) a set of tools for plotting and using these UV backgrounds to create ionization tables for analysis tools such as Trident \citep{trident-2017ApJ...847...59H}.

\begin{figure*}
    \centering
    \includegraphics[scale=0.55]{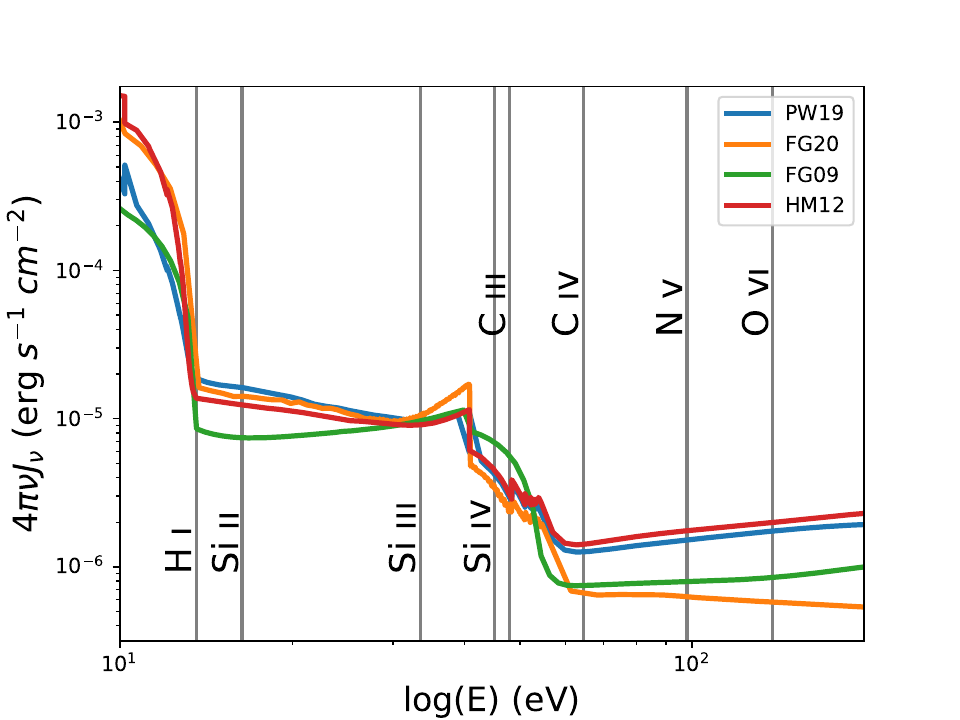} 
    \includegraphics[scale=0.5]{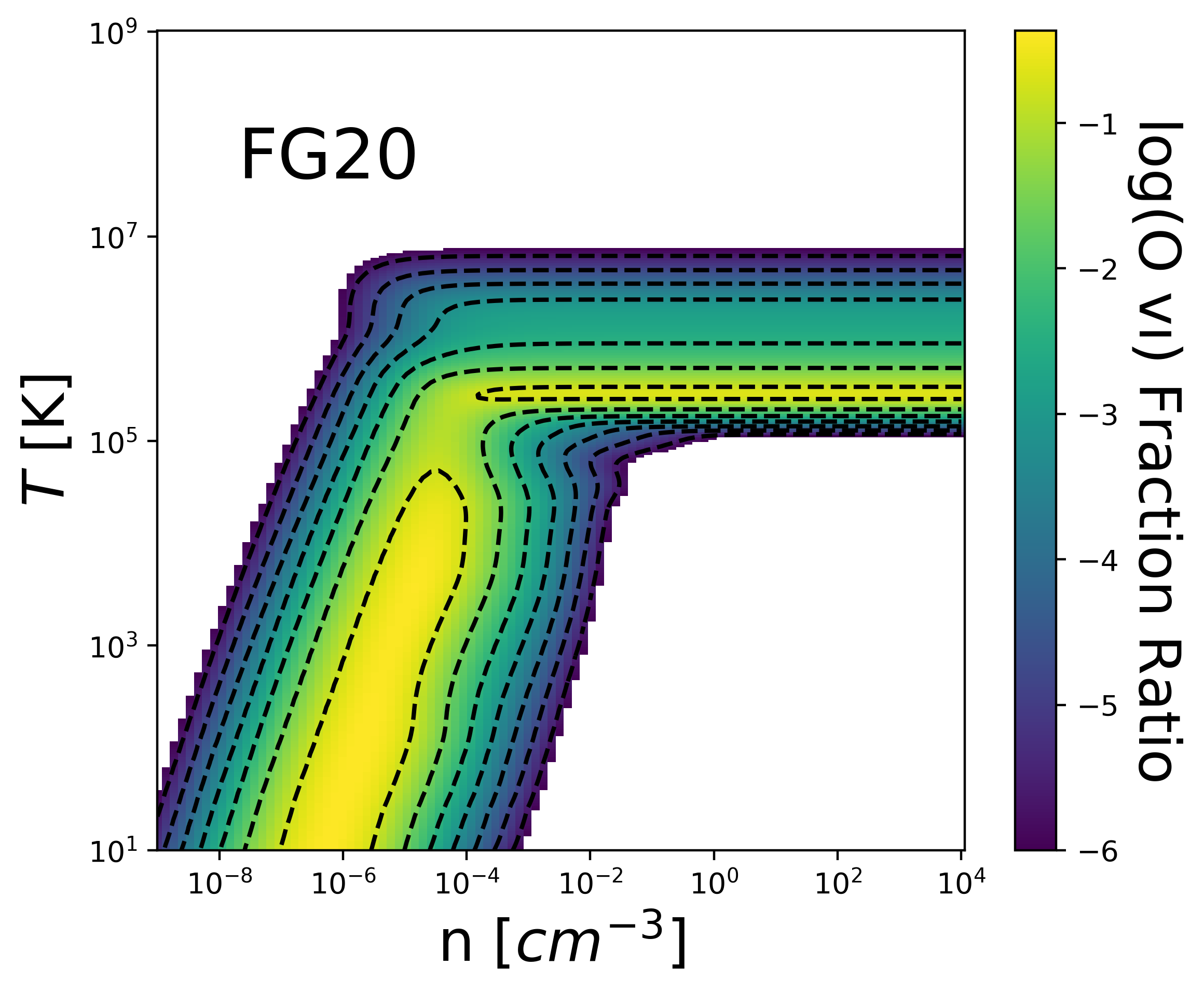}
    \caption{(Left) A plot displaying the intensity (in units of erg s$^{-1}$ cm$^{-2}$) of four different ultraviolet background models at z $\approx$ 2.5 as a function of photon energy (eV). Along the x-axis are vertical lines indicating the ionization energy of several ions that are commonly observed in the circumgalactic medium. (Right) A 2D histogram displaying the fraction of \ovi\ for the FG20 UVB model over a wide range of number densities and temperatures. In the x-axis is the log proper number density (cm$^{-3}$) of the gas while the y-axis has the log temperature (K).}
    \label{fig:thefigure}
\end{figure*}

\section{Data and Code Repository}

All of the code for creating these tables can be found at \dataset[Metagalactic\_Ultraviolet\_Background\_Data\_Tables]{https://doi.org/10.5281/zenodo.15367103}. This repository contains all of the consistently-formatted data tables along with the scripts used to generate them \citep{taira_2025}. It also contains code for running \texttt{cloudy\_cooling\_tools} (\url{https://github.com/brittonsmith/cloudy_cooling_tools}), a software package that uses the CLOUDY spectral synthesis code \citep{2013RMxAA..49..137F} to generate ionization tables with the data provided in this repository.

\subsection{UVB Tables}

The data tables are binned according to the intensity (in units of erg/s/cm$^2$) of the UVB at a given energy (in units of eV) and cover a wide range of redshifts which vary based on the data provided by the original authors. FG09 ranges from $z = 0.0 - 10.6$, HM12 from $z = 0.0 - 14.849$, PW19 from $z = 0.0 - 9.994$, and FG20 from $z= 0.0 - 10.0$.

A Python script named \texttt{make\_uvb\_tables.py} takes in UVB data formatted for CLOUDY and reformats the data into CSV files in intensity (erg/s/cm$^2$) binned in terms of energy (eV) for the range of redshifts. 

Additionally, there is a script included called \texttt{plot\_uvb\_spec.py}, which demonstrates 
a potential application of the code as seen in the left panel of Figure~\ref{fig:thefigure}.  This figure shows the intensity (in erg s$^{-1}$ cm$^{-2}$) of each UVB at $z \approx 2.5$ from 10 eV to $10^{2.3}$ eV on a logarithmic scale. Also plotted are the ionization energies of several notable CGM ions. 

\subsection{CLOUDY Functionality}

The software in this release also comes with some functionality to convert these CSV-formatted data tables back into CLOUDY format and input the data into the \texttt{cloudy\_cooling\_tools} 
package (URL provided above) to create tables of ionization fractions for different UVBs across a specified set of redshifts. 
This requires the user to install \texttt{cloudy\_cooling\_tools}.
The right panel of Figure \ref{fig:thefigure} shows a 2D histogram made with the FG20 UVB data for \ovi\ at $z=2.5$. The figure shows the log-scale fraction of oxygen that has been ionized to \ovi\ for the range of number densities and temperatures commonly used in galaxy formation. This figure can be recreated with any ion at any redshift using the plotting script included in this release, so long as those species and redshifts are included in the ionization table generated by CLOUDY and \texttt{cloudy\_cooling\_tools}.

\begin{acknowledgments}
The authors thank E.~Puchwein for sharing UV background data with us.
ET, CK, and BWO acknowledge support from NSF grants \#1908109 and \#2106575 and NASA ATP grants NNX15AP39G and 80NSSC18K1105. 
This work used resources provided by the Michigan State University High Performance Computing Center (operated by the MSU Institute for Cyber-Enabled Research). 
\end{acknowledgments}

\software{CLOUDY~\citep{2013RMxAA..49..137F}, \href{https://mpi4py.readthedocs.io/en/stable/mpi4py.html}{mpi4py}, \href{https://numpy.org/}{NumPy}, \href{https://scipy.org/}{SciPy}, \href{https://matplotlib.org/}{matplotlib}, \href{https://docs.h5py.org/en/stable/index.html}{h5py}, \href{https://pypi.org/project/roman/}{roman}}


\bibliography{sample631}{}
\bibliographystyle{aasjournal}

\end{document}